\documentclass[sigconf]{acmart}
\usepackage{array}
\usepackage{multirow}
\AtBeginDocument{%
  \providecommand\BibTeX{{%
    \normalfont B\kern-0.5em{\scshape i\kern-0.25em b}\kern-0.8em\TeX}}}

\copyrightyear{2021}
\acmYear{2021} 
\setcopyright{iw3c2w3}
\acmConference[WWW '21]{Proceedings of the Web Conference 2021}{April 19--23, 2021}{Ljubljana, Slovenia} 
\acmBooktitle{Proceedings of the Web Conference 2021 (WWW '21), April 19--23, 2021, Ljubljana, Slovenia}
\acmPrice{}
\acmDOI{10.1145/3442381.3450129}
\acmISBN{978-1-4503-8312-7/21/04}
\begin{document}

\title{Learning a Product Relevance Model from Click-Through Data in E-Commerce}


\author{Shaowei Yao}
\authornote{Equal contribution. This work was done when Shaowei Yao was an intern at Alibaba Group. Jiwei Tan is the corresponding author.}
\affiliation{%
  \institution{Peking University}
  \city{Beijing}
  \country{China}}
\email{yaosw@pku.edu.cn}

\author{Jiwei Tan}
\authornotemark[1]
\affiliation{%
  \institution{Alibaba Group}
  \city{Hangzhou}
  \country{China}}
\email{jiwei.tjw@alibaba-inc.com}

\author{Xi Chen, Keping Yang}
\affiliation{%
  \institution{Alibaba Group}
  \city{Hangzhou}
  \country{China}}
\email{{gongda.cx, shaoyao}@taobao.com}

\author{Rong Xiao}
\affiliation{%
  \institution{Alibaba Group}
  \city{Hangzhou}
  \country{China}}
\email{xiaorong.xr@taobao.com}

\author{Hongbo Deng}
\affiliation{%
  \institution{Alibaba Group}
  \city{Hangzhou}
  \country{China}}
\email{dhb167148@alibaba-inc.com}

\author{Xiaojun Wan}
\affiliation{%
  \institution{Peking University}
  \city{Beijing}
  \country{China}
}
\email{wanxiaojun@pku.edu.cn}

\renewcommand{\shortauthors}{Yao and Tan, et al.}

\begin{abstract}
  The search engine plays a fundamental role in online e-commerce systems, to help users find the products they want from the massive product collections. Relevance is an essential requirement for e-commerce search, since showing products that do not match search query intent will degrade user experience. With the existence of vocabulary gap between user language of queries and seller language of products, measuring semantic relevance is necessary and neural networks are engaged to address this task. However, semantic relevance is different from click-through rate prediction in that no direct training signal is available. Most previous attempts learn relevance models from user click-through data that are cheap and abundant. Unfortunately, click behavior is noisy and misleading, which is affected by not only relevance but also factors including price, image and attractive titles. Therefore, it is challenging but valuable to learn relevance models from click-through data. In this paper, we propose a new relevance learning framework that concentrates on how to train a relevance model from the weak supervision of click-through data. Different from previous efforts that treat samples as either relevant or irrelevant, we construct more fine-grained samples for training. We propose a novel way to consider samples of different relevance confidence, and come up with a new training objective to learn a robust relevance model with desirable score distribution. The proposed model is evaluated on offline annotated data and online A/B testing, and it achieves both promising performance and high computational efficiency. The model has already been deployed online, serving the search traffic of Taobao for over a year.
\end{abstract}





\maketitle

\section{Introduction}
In recent years online shopping has become indispensable in our life. Most of the customers start their online shopping journey by entering a query to describe the product they want. Nevertheless, commercial e-commerce search engines are usually trained to optimize for user's engagement and conversion, possibly at the cost of relevance in some cases \cite{DBLP:conf/wsdm/CarmelHLLM20}. Showing products that do not match search query intent undoubtedly degrades customer experience and possibly hampers customers' long-term trust and engagement. Therefore, it is a fundamental component of the e-commerce search engine to accurately judge whether the candidate products are relevant to the query.

In the e-commerce scenario, the queries are usually described by users in daily language while the products are described by sellers in professional language, where severe vocabulary gap may exist \cite{10.1145/3289600.3291039}. Thus, semantic matching is needed where the methods rely on hand-crafted features may not accurately address the relevance task. 
Neural models for IR typically accept the raw text of a query and document as input \cite{Mitra2017NeuralMF}, which perform well for semantic matching. An effective neural model usually requires sufficient, sometimes massive amounts of, data to train. Although manual annotation ensures the high quality of training data, it is expensive and time-consuming when the dataset is large. To address this problem, relevance models in Web search usually choose to train the model using the easy-to-get click-through data, and evaluate the model on the manually annotated evaluation/test sets. However, there are three problems that make learning relevance model in e-commerce challenging: 1) Different from Web search where user click can be viewed as the proxy of relevance, the click in e-commerce is much more noisy and misleading, which is not only affected by relevance, but also by many factors including price, attractive titles or images \cite{Zhang2019ImprovingSM}. 2) Un-clicked products do not necessarily mean irrelevant to the query. In a practical search engine, top candidates in a query session are usually at least somewhat relevant. The online system will rarely see completely unrelated candidates, which causes a serious distributional skew during serving \cite{10.1145/3326937.3341259}. It results in the trouble of introducing negative samples for training the relevance model. 3) Although successful attempts find that instead of training a binary classification model with $0/1$ labels, pair-wise training that predicts a positive candidate with a higher score than a negative candidate is helpful for mitigating the negative effect of the bias in click-through data \cite{10.1145/3326937.3341259,10.1145/3289600.3291039}, pair-wise models have the problem that they are not well calibrated to an absolute notion of relevance. Since the model only learns a product is better than another under a query, a score of $0.3$ can either be \textit{Good} or \textit{Bad} depending on the query, which brings problems when used to filter irrelevant candidates online.

In this paper, to address the above problems, we propose a new framework for e-commerce relevance learning. Our work concentrates on how to train an effective relevance model from e-commerce user click-through data. We not only propose a new relevance model, but also introduce a novel way to construct training data from search logs, and come up with a new training objective to learn a robust relevance model with desirable score distribution. 
Different from previous researches that treat samples as either relevant or irrelevant, we consider more fine-grained samples for training. We construct five types of training data according to their confidence of relevance category. Although clicked products are not guaranteed to be relevant to the query, the click behavior usually reflects the relative relevance in the way that higher Click-Through Rate (CTR) products will mostly be relevant to the query, while lower CTR products are possibly less relevant to the query. This inspires us that the CTR level can be the indicator of the relevance. Accordingly, we classify the clicked products into three positive categories (``strong relevant'', ``relevant'', ``weak relevant'') according to their CTRs, and hope that when training the model the relevance can be differentiated w.r.t. the CTR information. Since we consider the less clicked products as weak relevant instead of irrelevant, one kind of negative samples available can be random samples for the query. Unfortunately, these ``strong irrelevant'' samples are too easy for the model to distinguish, while in practical serving the irrelevant cases are much more difficult. To imitate real online negative samples, we introduce a query rewriting technique to generate ``weak irrelevant'' samples. By using the rewritten queries that are similar to the original queries but with different meanings (i.e. the rewritten queries with low confidence), the products related to low-confident rewritten queries can be seen as the ``weak irrelevant'' training instances of the original query. In order to better leverage the constructed fine-grained training data, we design a novel loss function with an inbuilt threshold to differentiate these different types of training samples. In this way, the model trained on this carefully constructed dataset can learn the relative relevance among candidates. 
Therefore, the proposed method is similar to pair-wise approaches but uses a point-wise loss function, which can not only mitigate the negative effect of the bias in the click-through data, but also learn a desirable output score distribution.
To summarize, the contribution of this paper is three-fold:
\begin{itemize}
\item Since the acquisition of training positive and negative samples is the dominant problem for e-commerce relevance learning, we propose a fine-grained sample construction method that can learn a robust semantic relevance model. Using the statistical CTR information for training samples also alleviates the noise of a single click behavior.
\item Although pair-wise training usually achieves better performance, it suffers from unexpected score distribution and explosion of training pairs. We propose a novel training paradigm that can maintain the advantage of pair-wise training but in a point-wise fashion.
\item The proposed model is evaluated both on offline human-annotated datasets in Chinese and online A/B testing, achieving promising performance as well as high computation efficiency. The model has already been deployed online and served the entire search traffic of Taobao\footnote{\url{https://www.taobao.com/}}, the largest Chinese e-commerce portal for over a year.
\end{itemize}
The rest of this paper is organized as follows. In Section 2 we introduce related work. The proposed method is detailed in Section 3, and experimental results are presented in Section 4. Finally, in Section 5 we conclude this work.

\section{Related Work}
\subsection{Text Matching}
Text matching is a long-standing problem, where the model takes two textual sequences as input and predicts a numerical value or a category indicating their relationship. 
Early work mostly performs keyword-based matching relies on manually defined features, such as TF-IDF similarity and BM25 \cite{robertson1995okapi}. These methods cannot effectively utilize raw text features and usually fail to evaluate semantic relevance. Recently neural-based text matching has achieved promising performance.  These neural-based methods can be roughly divided into three categories. 
First is \textit{representation-based} models that individually encode both the query and the document into single embedding vectors. 
DSSM \cite{huang2013learning} is one representative architecture that employs two separate deep fully-connected networks to encode the query and the document. Meanwhile, more sophisticated architectures can be adopted to enhance the ability of learning semantic representations. For example, ARC-I \cite{10.5555/2969033.2969055} and CDSSM \cite{10.1145/2661829.2661935} use CNNs to adequately model the internal structures of language objects and the interaction between them. LSTM-DSSM \cite{palangi2014semantic} and LSTM-RNN \cite{10.1109/TASLP.2016.2520371} use RNNs to explicitly model word dependencies in the sentences. Typically dot-product, cosine, or parameterized non-linear layers are used to measure the similarity between query and document representations. Since the embeddings of queries and documents can be pre-computed offline, representation-based methods are online efficient and therefore are widely used in industrial search engines. However, the encoding procedure of two inputs is independent with each other, making the final classifier hard to predict their relationship.

The second line of research in this area is \textit{interaction-based} models which solve the above problems. The interaction-based models first match different parts of the query with different parts of the document at low level, and then aggregate the partial evidence of relevance to make the final decision. Sophisticated techniques can be introduced in the aggregation procedure.  ARC-II \cite{10.5555/2969033.2969055} and MatchPyramid \cite{10.5555/3016100.3016292} use CNN to learn rich hierarchical matching patterns over the matching matrix. 
Match-SRNN \cite{10.5555/3060832.3061030} further models the recursive matching structure to better capture long-distance dependency between the interactions. DecompAtt \cite{parikh-etal-2016-decomposable} leverages attention mechanism for alignment. These methods capture more interactive features between inputs so they bring significant improvement. But the interaction-based models are mostly time-consuming which is hard to be deployed in practical online service.

More recent studies are built upon pre-trained language models. The most notable example is BERT \cite{devlin-etal-2019-bert}, a pre-trained deep bidirectional transformers model. The typical paradigm of BERT-based relevance learning is to feed the query-document pair into BERT and then build a non-linear classifier upon BERT's [CLS] output token to predict the relevance score \cite{nogueira2019passage}. 
\citeauthor{nogueira2019multistage}~\shortcite{nogueira2019multistage} propose duoBERT that learns the relevance of a pair of texts in a pair-wise fashion. With extremely large corpus for pre-training, these methods can achieve new state-of-the-art performance on various benchmarks but are highly expensive in practice.

The e-commerce relevance task can also be viewed as a text matching problem. Unfortunately, practical online systems require high concurrency and low latency, making the interaction-based or more sophisticated models difficult to be deployed online. Therefore, in this work our model follows the representation-based architecture. Nevertheless, in this work we enhance the representation model by proposing more fine-grained interaction with multi-aspect attention to compute multiple latent vectors of query and documents followed by the interaction layer to integrate them. This strategy can yield richer interaction between the query and
products and outperforms strong representation-based models. It still maintains the advantages of representation-based methods and is highly efficient for online serving.

\subsection{Web Search Relevance Learning}
The related researches we describe in the previous section are mainly focused on text matching, where two input texts are semantically similar and homogeneous (have comparable lengths). In the area of Web search, the input is usually a query-document pair. The query and document are very different in lengths, and the query may match only part of the document. Typical Web search relevance learning methods can be categorized into two classes: based on global distribution of matching strengths and based on local context of match terms \cite{10.1145/3289600.3291380}. The global distribution based models first calculate the matching signals among the document for each query and then aggregate the distribution. For example, DRMM \cite{10.1145/2983323.2983769} introduces pyramid pooling to summarize the word-level similarities, where the histogram counts the number of word pairs at different similarity levels. K-NRM \cite{10.1145/3077136.3080809} uses kernel-pooling to summarize the translation matrix and provide soft match signals. Conv-KNRM \cite{10.1145/3159652.3159659} further uses CNN to represent n-grams of various lengths and softly match them in a unified embedding space. While matching strength distributions is robust compared with raw matching signals, it loses the term order information when calculating the distributions.

The local context based models first identify the local context of each query among document and conduct matching between query and local context, and then aggregate the local matching signals for relevance learning. DeepRank \cite{10.1145/3132847.3132914} is one of the notable local context based models. It extracts the relevant contexts and determines local relevance by utilizing CNN or 2D-GRU. Finally, an aggregation network with sequential integration and term gating mechanism is used to produce a global relevance score. These local context based models filter out the irrelevant parts of documents so that improve robustness. Meanwhile, the sequential integration keeps the order information within each local context. 

The Web search relevance task is similar to e-commerce relevance that it also models the semantic matching between query and document, which suffers from vocabulary gap. Both tasks require high efficiency therefore representation-based architecture is more favorable. Differently, in Web search the query and document are usually very different in length, making most methods not feasible for the e-commerce relevance task. Both tasks have abundant click-through data, and usually learn relevance models from the click-through data. Unfortunately, different from Web search that user clicks can be viewed as the proxy of relevance, the click in e-commerce is much more noisy and misleading. In this work, to mitigate the negative effect of the bias in e-commerce click-through data, we proposed a novel data-construction method to learn a robust relevance model from the more noisy click-through data.

\subsection{E-commerce relevance learning}
E-commerce relevance learning is an important part of the product search. Currently there is not 
a commonly-used public benchmark for e-commerce relevance, so previous works usually evaluate their models on the online service and the real-world dataset constructed by themselves. In addition, most previous semantic relevance models use click-through data as implicit feedback label. For example, \citeauthor{10.1145/3326937.3341259}~ \shortcite{10.1145/3326937.3341259} propose a typical framework for e-commerce relevance learning. A Siamese network is adopted to learn pair-wise relevance of two products to a query. They investigate training the model with user clicks and batch negatives, followed by fine-tuning with human supervision to calibrate the score by pair-wise learning. This framework is taken as the baseline of our model. \citeauthor{10.1145/3289600.3291039}~\shortcite{10.1145/3289600.3291039} propose a co-training framework to address the data sparseness problem by investigating the instinctive connection between query rewriting and semantic matching. The idea of leveraging query rewritten candidates inspires us with a brilliant way to produce difficult irrelevant samples. 
\citeauthor{Zhang2019ImprovingSM}~\shortcite{Zhang2019ImprovingSM} also find the weakness of training with click signals, and address this problem by proposing a multi-task learning framework of query intent classification and semantic textual similarity to improve semantic matching efficiency.
Recently \citeauthor{10.1145/3292500.3330759}~\shortcite{10.1145/3292500.3330759} 
introduce a $3$-part hinge loss to differentiate multiple types of training data. They classified training instances into three categories: random negative examples, impressed but not purchased examples, and purchased items. Nevertheless, this way of classifying data is a bit rough and the purchase is pretty sparse, which cannot provide enough supervision signals. Meanwhile, they do not consider the bias and noise in the click-through data. 
Their idea is relevant to this work, but different from them we propose a more fine-grained data construction method and correspondingly introduce a new training objective to learn a robust relevance model with desirable score distribution.
  
\section{Methodology}
In this section, we will first introduce how the training dataset is constructed from the click-through data. Then we will describe the model architecture. Finally, we will present a new paradigm for training the relevance model.

\subsection{Data Construction} \label{data_construction}
Most previous researches use the easy-to-get click-through data for training the relevance model. For example, the products which are clicked or purchased with a query are usually viewed as positive instances, while products that are not clicked with the query are treated as negative instances. 
However, the relevance model is to be evaluated on the human-annotated dataset, which is different from the click-through data. As we mentioned above, the click behavior can be affected by many other factors besides relevance. In our practice, this discrepancy of training and test dataset results in that the model performance is very sensitive to the distribution of training data. Previous studies find training a pair-wise model will be better than learning the 0/1 label with point-wise loss \cite{10.1145/3326937.3341259,10.1145/3289600.3291039}. One assumption by us is that relevance can be learned through learning comparison of products since clicked products are possibly more relevant than not clicked ones; this is more reasonable than learning an absolute relevant or irrelevant label since the click labels do not equal relevance. 

This inspires us that for effective training of a relevance model, there should be more reasonable relevance samples and learning targets. Hence, we propose to construct a more fine-grained training dataset to depict subtle relevance difference. Our constructed dataset contains five types of training instances, including three types of positive instances and two types of negative instances, corresponding to ``strong relevant'', ``relevant'', ``weak relevant'', ``weak irrelevant'' and ``strong irrelevant'', respectively. Each type of training data has a specially designed threshold in the loss function to help the model differentiate it. It will be detailed in Section \ref{training}. When constructing relevance instances from the click-through data, a key point is to eliminate the position bias so that click behavior will be more related to relevance. 

\begin{figure}[t]
    \centering
    \includegraphics[width=0.9\columnwidth]{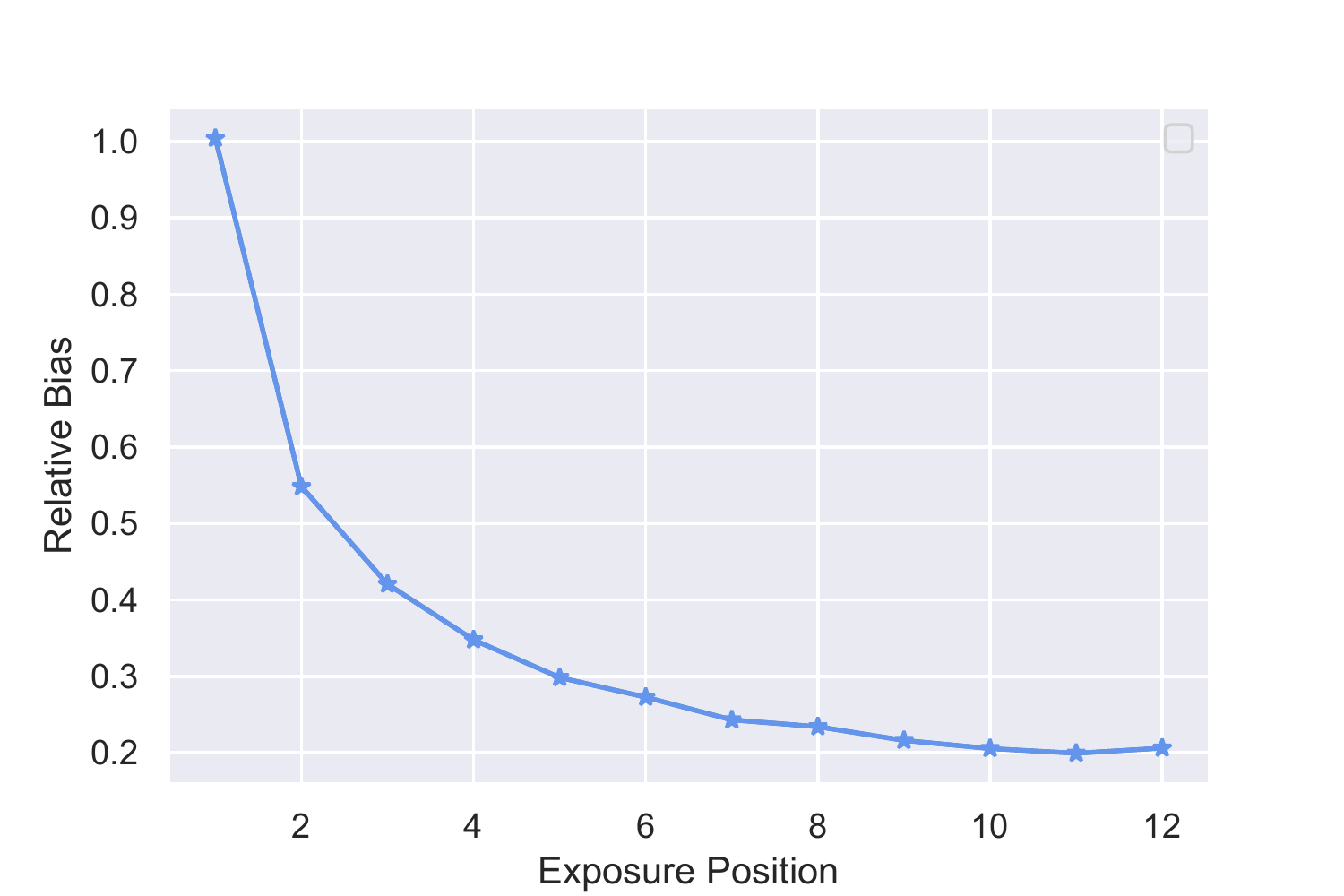}
    \caption{The relative estimated position bias in terms of the first position. The top-ranked results are likely to get more clicks than the results ranked lower. Thus there is usually a higher bias in the front position. }
    \label{position_bais}
\end{figure}

\noindent {\bf Position bias estimation}: User click is strongly influenced by the exposure position of products, and how to eliminate the influence of {\em position bias} has been previously studied \cite{10.1145/3018661.3018699}. 
In this work,  we will first eliminate the position bias before using the CTR statistic feature. 
To realize this, we conduct a well-designed online experiment to estimate position bias. Specifically, for each of all the online queries, the products presented in the first page will be randomly re-ranked. As a result, the statistical CTR of a query with all products will equal its real CTR (CTR without the influence of position), because in this case each product has the same probability to appear at each position under the query, and hence no position bias is incorporated when calculating the CTR. Formally, the real CTR of a query with all products $ctr_{\text{real}}$  just equals:

\begin{equation}
    ctr_{\text{real}} = \frac{click_{\text{overall}}}{exposure_{\text{overall}}}
\end{equation}
where $click_{\text{overall}}$ and $exposure_{\text{overall}}$ represent the total times of click and exposure for all products under a query, respectively. After getting the real CTR for a query, the position bias at position $i$ can be derived from the difference between the statistical CTR at position $i$ and the overall real CTR. More formally, at each position $i$, we calculate the statistical CTR as:

\begin{equation}
% \begin{aligned}
     ctr^i_{\text{stat}} = \frac{click_i}{exposure_i}
% \end{aligned}
\end{equation}
where $click_i$ and $exposure_i$ denotes the times of click and exposure for all products under the query at position $i$, respectively. Then, the position bias $bias_i$ will be the ratio of statistical CTR $ctr^i_{\text{stat}}$ to the real CTR  $ctr_{\text{real}}$:

\begin{equation}
% \begin{aligned}
     bias_i = \frac{ctr^i_{\text{stat}}}{ ctr_{\text{real}}}
% \end{aligned}
\end{equation}

The overall position bias at each position is the average of $bias_i $ for all queries. Figure \ref{position_bais} depicts the relative bias at each position compared to the first position.

\noindent {\bf Positive instances}: We generate positive instances from the clicked products of a query. However, irrelevant products can also acquire clicks although their CTRs are probably lower. Therefore, the CTR can be an indicator of relevance in the way that higher CTR products will mostly be relevant while lower CTR products may have problems in relevance (and there are also other reasons possibly). To make the CTR feature more reliable, we only take the exposure in the first page into statistics, and incorporate the position bias described above to make the CTR more related to relevance. Specifically, because in real online traffic each query-product pair is not guaranteed to appear evenly at each position, the statistical CTR will be first divided by the estimated position bias to eliminate the position effect and get the calibrated CTR.  Then we divide the products under a query according to the calibrated CTR. The instances with the calibrated CTR ranking in top 20\%, tail 20\%, and others are categorized into ``strong relevant'', ``weak relevant'' and `` relevant'' instances, respectively.    

\noindent {\bf Negative instances}: Considering negative instances, the irrelevant products for a query can be easily obtained in a way of random sampling from the product pool. Usually such negative instances can be easily distinguished by the model, and therefore are of little help for the model to learn an accurate decision boundary to decide relevant or not. In online serving the products returned by retrieval process are mostly relevant to the query, so the model has to distinguish the hard-to-classify negatives. To this end, we construct the ``weak irrelevant'' instances which will be hard to classify. Inspired by \citeauthor{10.1145/3289600.3291039}~\shortcite{10.1145/3289600.3291039}, we employ the query rewriting (QR) technique  to generate hard-to-classify negative instances. Specifically, a GBDT-based online QR model will produce the candidate rewritten queries and their confidence scores. Most rewritten queries with low confidence change the intent of the original query. Thus the products clicked under these rewritten queries are often irrelevant to the original query. For example, given a query ``red dress'', a  low-confidence rewritten query can be ``white dress''. The products clicked under the query ``white dress'' will be probably irrelevant to ``red dress''. The negative instances generated in this way will be difficult for a relevance model to distinguish, and are very similar to online irrelevant cases. These negative training instances can considerably improve the distinguishing ability of the model. The threshold of low confidence is empirically chosen to make more than 90\% of the produced products really irrelevant to the original queries.

\subsection{Model Architecture}
\begin{figure*}[tb]
    \centering
    \includegraphics[width=0.8\linewidth]{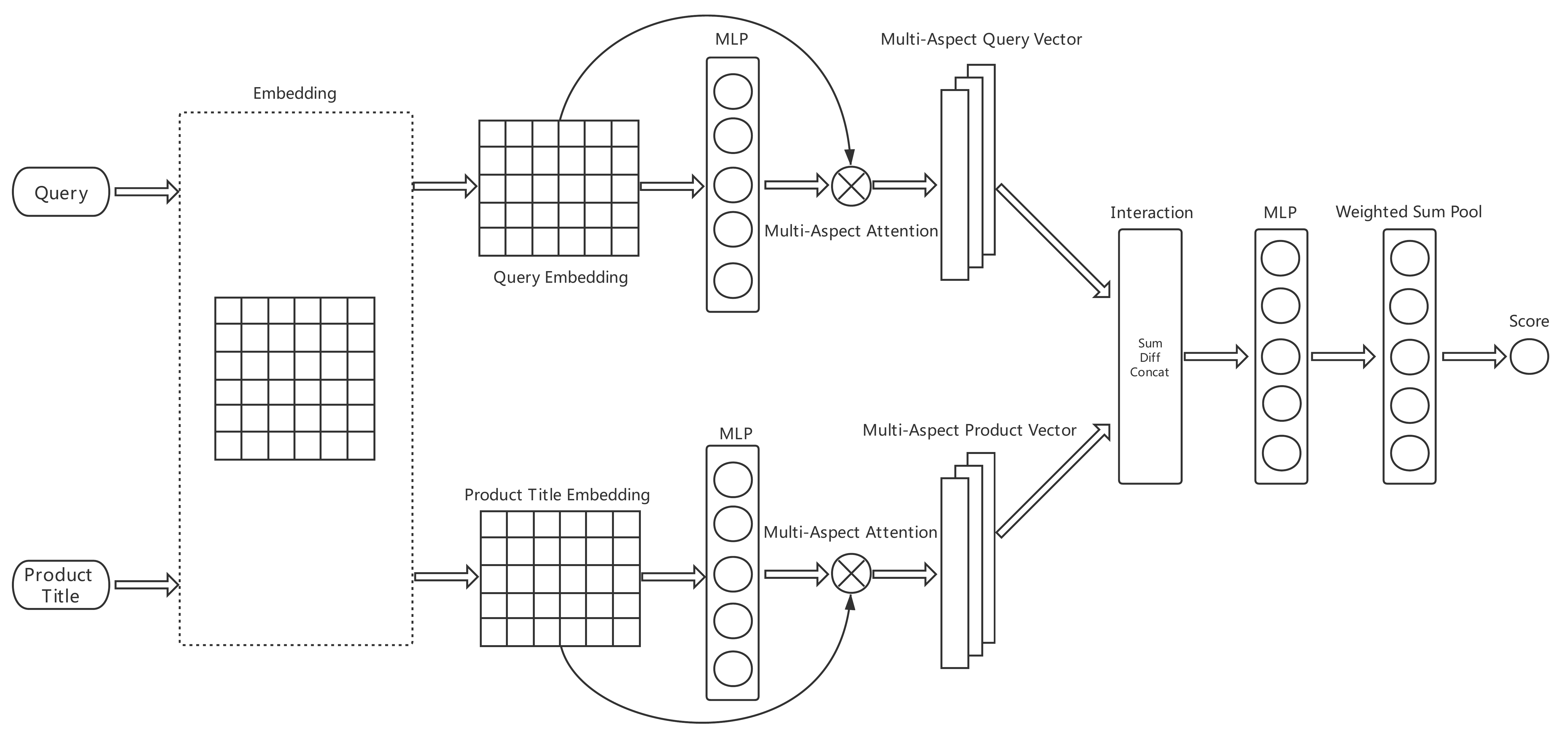}
    \caption{The architecture of our proposed model MASM. The query and product encoders have the same architecture which consists of 1) a shared embedding layer; 2) a MLP layer; 3) a multi-aspect attention layer. The outputs of query vectors and product vectors are combined via an interaction layer to get the matching representation, which is finally used to predict the final matching score. }
    \label{architecture}
\end{figure*}

Our model, named as Multi-Aspect Semantic Model (MASM), is illustrated in Figure \ref{architecture}.  The inputs are the word sequences of a query and the title of a product, and the output will be a relevance score. Considering online efficiency, we adopt the representation-based architecture instead of interaction-based architecture. A common practice in representation-based methods is mapping the query embeddings into a single latent vector according to calculated attention weights. The main benefit of this strategy is that the latent vector can be computed offline to speed up online query processing. Unfortunately, compressing the information of the query or product into a single latent vector independently will lose much fine-grained matching information, which is a critical reason why representation-based methods usually perform worse than interaction-based methods. But the interaction-based methods suffer from the expensive computational complexity which is hard to be deployed in online services. To make richer interaction between the query and products but without significantly increasing computational costs, we propose the {\em multi-aspect attention} to compute multiple latent vectors of the query as well as products in different semantic spaces, then an interaction layer integrates these latent vectors and computes the matching representation. This strategy can yield richer interaction between the query and products while maintaining the advantages of representation-based methods. Formally, we denote the embeddings of two input sequences as %$q = (q_1, q_2, ..., q_n)$ of $n$ elements and $t = (t_1, t_2, ..., t_m)$ of $m$ elements 
$\boldsymbol{e_q}$ (query) and $\boldsymbol{e_p}$ (product title), where $ \boldsymbol{e_q} \in \mathbb{R}^{n\times d}$ and $\boldsymbol{e_p} \in \mathbb{R}^{m\times d}$. $d$ is the embedding dimension and $n, m$ are the length of query and product title, respectively. The multi-aspect attention will be performed over projected embeddings of two sequences to get the multi-aspect vectors. Taking query $\boldsymbol e_q$ as an instance, the projected query representation $\boldsymbol{We_q} $ performs self-attention and then passes through a ReLU activation function to get the self-attention weights $\boldsymbol{\hat{\alpha}} \in \mathbb{R}^{n\times n}$. The convolution operation with $h$ kernels calculates multi-aspect  attention weights $\boldsymbol \alpha = (\boldsymbol{\alpha}_1, \boldsymbol{\alpha}_2, ..., \boldsymbol{\alpha}_h)$. Then the multi-aspect query vectors $\boldsymbol q = (\boldsymbol{q}_1, \boldsymbol{q}_2, ..., \boldsymbol{q}_h)$ are computed according to $\boldsymbol \alpha$. 

\begin{equation}
    \begin{aligned}
        \boldsymbol{\hat{\alpha}}  &= \sigma \left( \boldsymbol{W_\text{Q}e_q} \left( \boldsymbol{W_\text{K}e_q}\right)^T \right) \\
        \boldsymbol \alpha &= (\boldsymbol{\alpha_1, \alpha_2, ..., \alpha_h}) = \text{Conv1D}\left( \boldsymbol{\hat{\alpha}} \right) \\
        \boldsymbol q &= (\boldsymbol{q_1, q_2, ..., q_h}) = \boldsymbol \alpha \cdot \boldsymbol{e_q} \\
    \end{aligned}
\end{equation}
where $\boldsymbol{\alpha}_i \in \mathbb{R}^{1\times n}$, and $\boldsymbol{q}_i  \in \mathbb{R}^{d}$. $\boldsymbol{W_\text{Q}, W_\text{K}}$ are trainable parameter matrices of self-attention.
The multi-aspect product vectors $\boldsymbol p = (\boldsymbol p_1, \boldsymbol p_2, ..., \boldsymbol p_h)$ are calculated in the same way. Then these vectors are combined via Eq.~\ref{combine} (including summation, subtraction and concatenation) to get the interacting representation $\boldsymbol r $, which indicates the multiple matching representations of query and the product: 
\begin{equation} \label{combine}
\begin{aligned}
    \boldsymbol r &= (\boldsymbol{r}_1, \boldsymbol{r}_2, ...,\boldsymbol r_h) \\
    \boldsymbol r_i &= \text{Concat} \left(\boldsymbol q_i, \boldsymbol p_i, \boldsymbol q_i+ \boldsymbol p_i, \boldsymbol q_i- \boldsymbol p_i \right) 
\end{aligned}
\end{equation}
where $\boldsymbol r_i \in \mathbb{R}^{4d}$. Finally the MLP (Multi-Layer Perceptron) projects multiple matching representations $\boldsymbol r$ into score vector $\boldsymbol s \in \mathbb{R}^{h}$ and performs weighted sum pooling to get the final matching score $s_{\text{final}}$ according to Eq.~\ref{model_output}. The activation function for the first MLP is tanh and the activation function for the weighted sum pooling is sigmoid to make the output final score in range 0 to 1.

\begin{equation}
\begin{aligned} 
    & \boldsymbol s_\text{aspects} = (s_1, s_2, ..., s_h) = \text{MLP}(\boldsymbol r) \\
    & s_{\text{final}} = \text{WeightedSumPool} \left( s_1, s_2, ..., s_h \right) \\
\end{aligned}
\label{model_output}
\end{equation}

\subsection{Model Training} \label{training}
\begin{table}[tbp]
    \centering
    % \scalebox{0.85}{
    \begin{tabular}{p{4cm}|p{3cm}<{\centering}}         
    \toprule
     \bf Data Type & \bf Threshold Value \\
     \midrule
     \midrule
     strong relevant & 0.9 \\
     relevant & 0.8 \\
     weak relevant & 0.6 \\
     \midrule
     weak irrelevant & 0.3  \\
     strong irrelevant & 0.1 \\
     \bottomrule
    \end{tabular}
    % }
    \caption{Particular threshold values for different types of training instances. These thresholds are built in the loss function to map the matching scores of different types of data into desirable score distribution intervals.}
    \label{threshold}
\end{table}
As described above, our training set is fine-grained, containing five types of training instances different from their relative relevance. To better leverage the dataset, we introduce a novel point-wise loss to make our model learn the difference between training instances, which keeps the advantage of pair-wise training. Specifically, for each type of training pair $(q, p, \text{type})$, we design a particular threshold $t_{q, p}$ as presented in Table \ref{threshold}. The thresholds are empirically designed, to meet our requirement for the online score distribution: 1) Strong relevant instances should have scores close to 1.0, while strong irrelevant instances should have scores close to 0.0; 2) Relevant instances should have scores larger than 0.5, and irrelevant instances should have scores smaller than 0.5. Then, the loss function is formulated as follows:

\begin{equation}
    \mathcal{L} = \sum_{q}\sum_{p}\max\left(\text{sign}\left(t_{q,p}-0.5\right)\times\left(t_{q,p}-s_{q,p}\right),0\right)
\end{equation}
where $s_{q,p}$ represents the predicted match score, i.e., $s_{\text{final}}$ in Eq.~\ref{model_output} for query $q$ and product $p$. The intuitive explanation of the loss function is to make the predicted score of a positive instance be greater than its corresponding threshold. Relatively, the predicted score of a negative instance should be smaller than its corresponding threshold. In this way, the relative relevance between different instances can be learned by the model in a point-wise fashion. Moreover, the threshold is a soft label, in that the loss will equal 0 as long as the score is higher than the threshold for a positive instance and vice versa. The soft loss has the advantage that will be robust to noisy training data. For example, the model can also predict a ``weak relevant'' instance to a score close to 1 when necessary and no penalty is introduced in the loss. Such case is widely observed like an expensive product may have lower CTR but is still perfect relevant to the query. Moreover, the threshold value can be set empirically instead of elaborately chosen owing to the soft loss, and the model performance shows stable in our practice when the thresholds are chosen in a reasonable range, i.e. ``strong relevant'' is higher than ``relevant'' and higher than ``weak relevant relevant'', and are all higher than 0.5; ``weak irrelevant'' has higher threshold than ``strong irrelevant'' and are both less than 0.5. 
 As a result, the predicted scores of the positive samples are usually close to $1$, and the predicted scores of the negative samples are usually close to 0. Those difficult samples will get mid scores representing the confidence of the model. Therefore, optimizing this point-wise loss can learn a relevance model with desirable score distribution, while pair-wise loss fails to guarantee the magnitude of predicted scores.
\subsection{Fine-tuning} \label{finetune}
Till now, the proposed method does not involve manually labeled data at all. In practice there are usually more or less labeled data available, and it is desirable the model can be further improved with labeled data. Fine-tuning is such an effective way to improve the model performance with high-quality training data. In this section we show that our method can also work well with labeled data and achieve further improvement in fine-tuning.

Human annotation is the gold-standard for the relevance task, and the label is set to 1 or 0 representing \textit{relevant} or \textit{irrelevant}, respectively. When fine-tuning, the model is trained with the Mean Square Error (MSE) loss to minimize the gap between the label $l_{q,p}$ and the output score $s_{q,p}$ of our model:

\begin{equation}
    \mathcal{L}_{MSE} = \frac{1}{n} \sum_{q}\sum_{p}\left\|l_{q,p}-s_{q,p}\right\|^{2}
\end{equation}
where $n$ is the total number of training samples. 

\section{Experiments}
\subsection{Dataset} \label{dataset}
 Our training dataset is collected from the search logs of Taobao in a year, and constructed as described in Section \ref{data_construction}. We have a total training samples of 6.2 billion, with the statistics of the dataset detailed in Table \ref{clickthrough_data}.  
\begin{table}
\begin{tabular}{c|ccc}
\toprule
\bf Data Type & \bf \#sample & \bf \#query & \bf \#product\tabularnewline
\midrule
\midrule
Strong relevant & 548,946,058 & 13,039,303 & 40,785,899\tabularnewline
Relevant & 653,660,431 & 11,645,344 & 66,465,899\tabularnewline
Weak relevant & 82,702,879 & 1,033,369 & 24,535,235\tabularnewline
\midrule
Weak irrelevant & 3,656,998,276 & 3,979,174 & 19,167,446\tabularnewline
Strong irrelevant & 1,282,345,887 & 13,042,775 & 78,900,274\tabularnewline
\midrule
All & 6,224,653,531 & 13,042,775 & 79,841,052\tabularnewline
\bottomrule
\end{tabular}
\caption{Statistics of the constructed training dataset.}
\label{clickthrough_data}
\end{table}

We also have a human-annotated dataset for the evaluation. The query-product pairs are first sampled from the search logs, and then labeled \textit{Good} (relevant) or \textit{Bad} (irrelevant) by experienced human annotators. This is a daily task running in Taobao, which has accumulated more than one million labeled samples. The dataset is split to training, validation and test set, as detailed in Table \ref{human_data}. The amount of annotation data is still limited to train a model from scratch. Therefore, the training set is typically used in fine-tuning our model trained from the constructed data, in order to compare with those state-of-the-art models trained or fine-tuned from annotation data. The validation set is used for hyper-parameter tuning and early-stopping of training. 

To further compare with state-of-the-art e-commerce relevance models, we also evaluate our model on another test set used by \citeauthor{10.1145/3289600.3291039}~\cite{10.1145/3289600.3291039}. This is also an e-commerce semantic relevance dataset labeled by human annotators. The test set is provided by the authors of \cite{10.1145/3289600.3291039} with the scores of their methods for our comparison. But we do not have the original training set, therefore we directly test our models on this test set. Our annotation data for fine-tuning are verified that do not overlap with this test set. 
\begin{table}
\begin{tabular}{c|ccccc}
\toprule
\bf Dataset & \bf \#sample & \bf \#query & \bf \#product & \bf \#good & \bf \#bad \tabularnewline
\midrule
\midrule
Train & 869,231 & 79,121 & 740,406 & 704,017 & 165,214\tabularnewline

Valid & 217,370 & 22,544 & 198,636 & 176,131 & 41,239\tabularnewline

Test & 217,307 & 22,618 & 198,556 & 175,527 & 41,780\tabularnewline
\bottomrule
\end{tabular}
\caption{Statistics of the human annotation dataset.}
\label{human_data}
\end{table}

\subsection{Experimental Setup} \label{setup}
The size of word vocabulary is 1,200,000 and the embedding dimension of queries and product titles are both $64$. The parameter size of the first MLP used to project the input embeddings is $(64 \times64)$. Note that the MLP used to project query embeddings is independent with the MLP used to project product embeddings. The number of convolution kernels in multi-aspect attention is $10$ and the dimension of multi-aspect query/product vector is also $64$. The last MLP projects the input vector into a scalar, thus its parameter size is $64\times 1$.

We use Tensorflow to implement our model and train the model using Adam optimizer. The hyper-parameters of Adam optimizer are $\beta_1 = 0.9, \beta_2=0.999, \epsilon=1e-8$ and the learning rate is set to $0.0001$. The batch size for training is set to 512. These hyper-parameters are chosen from the experimental results on the validation set. We train the model on a Tesla P100 GPU card and it usually takes about 3 days for the model to converge, after about 3 to 5 epochs. The convergence is reached when the ROC-AUC does not improve on the validation set.  

For fine-tuning, the learning rate is set to $0.0001$, and all the model parameters will be updated. We monitor the ROC-AUC on the validation set after each training epoch. The best results are usually achieved in about $10$ epochs. 

\subsection{Baselines and Evaluation Metrics}
The proposed model is named as \textit{MASM}. The constructed dataset from Taobao click-through data is named as {\em Level-Wise Relevance} (\textit{LWR}) dataset, and our method is named as {\bf MASM+LWR}. The model which is further fine-tuned with annotation data is named as {\bf MASM+LWR+Finetune}. For comparing with the proposed \textit{LWR} dataset, we introduce two baselines: 1) {\bf MASM+Finetune} which is to train the MASM model with the human annotated training data only, to verify the difficulty of training a product relevance model without large-scale click-through data; 2) {\bf MASM+Click} which is trained on the original click-through data with clicked samples as positive instances and un-clicked samples as negative instances, in a pair-wise training fashion like \citeauthor{10.1145/3326937.3341259}~ \shortcite{10.1145/3326937.3341259}. Note that when using the click-through data for training the baseline, point-wise loss fails to converge and therefore we do not report here.  In addition, we adopt several existing state-of-the-art methods for comparison as follows:

\begin{itemize}
    \item {\bf DSSM} \cite{huang2013learning}: The well-known representation-based model that employs two separate deep fully-connected networks to encode the query and the product title. DSSM has a similar architecture to our model so we take it as one of our baselines to verify the effectiveness of our model architecture. We use the DSSM model trained with Taobao click-through data, which is implement in AliNLP\footnote{\url{https://www.aliyun.com/product/bigdata/product/nlp}}.
    \item {\bf BM25 Sup (Rank+Embed)} \cite{10.1145/3077136.3080832}: 
    A deep neural ranking model of pair-wise paradigm, which is trained on about $5$ billions of training examples that are annotated by BM25 as the weak supervision signal. It uses a learnable representation function to compute the weighted sum of the word embedding as the input, which embeds the context information in each word embedding to help the model perform soft semantic matching. 
    \item {\bf Co-training} \cite{10.1145/3289600.3291039}: The latest state-of-the-art semantic relevance method in e-commerce search. This method utilizes a co-training framework built for query rewriting and semantic matching, where two models are first pre-trained using click-through data and then iteratively co-trained by the weakly labeled data generated for each other. The model is trained on three types of data, including human-labeled dataset, unlabeled triples set and click-through data. The sizes of these training data are about $1.4$M, $12$ billion and $1$ billion, respectively. 
    \item {\bf SBERT} \cite{reimers-gurevych-2019-sentence}
    : A strong baseline implemented by us which is inspired by \citeauthor{reimers-gurevych-2019-sentence}~\cite{reimers-gurevych-2019-sentence}. It is a representation-based architecture but uses pre-trained BERT \cite{devlin-etal-2019-bert} to produce the query and product embeddings. The query and the title are encoded independently using a shared BERT encoder. The final output score is computed by the cosine distance of the query embedding and the title embedding. We initialize the BERT from Google pre-trained weights\footnote{\url{https://storage.googleapis.com/bert_models/2018_11_03/chinese_L-12_H-768_A-12.zip}}, and then fine-tune the entire model with the human annotation data described in Table \ref{human_data}. The model is fine-tuned to achieve the best results on the validation set.
\end{itemize}

 For the evaluation of the product relevance task, human annotation is viewed as the ground truth, where  \textit{Good} and \textit{Bad} annotation are taken as 1/0 labels. Therefore, it can be viewed as a classification task, and the Area Under Curve (AUC) is widely used as the evaluation metric in e-commerce scenarios \cite{DBLP:conf/kdd/ZhouZSFZMYJLG18,10.1145/3326937.3341259}. Receiver Operator Characteristic (ROC) curve is the most commonly used for measuring the results of binary decision problems. Meanwhile, considering actual relevance datasets are usually highly skewed, we also use the Precision-Recall curve for evaluation, which is more informative in this scenario \cite{10.1145/1143844.1143874}. Note that in the e-commerce relevance scenario, most instances are positive and we are more concerned about negative instances. Therefore the PR-AUC used in this paper is the negative PR-AUC that treats \textit{Bad} as 1 and \textit{Good} as 0. We denote these two metrics as ROC-AUC and Neg PR-AUC in this paper.

\begin{table}[t]
    \centering
    % \begin{tabular}{p{4cm}|p{4cm}<{\centering}|cc}
    \scalebox{1.0}{
    % \begin{tabular}{l|m{1.5cm}<{\centering}|m{1cm}<{\centering}c}
    \begin{tabular}{l|m{2cm}<{\centering}|m{2cm}<{\centering}}
         \toprule
         \bf Model & \bf ROC-AUC & \bf Neg PR-AUC \\
         \midrule
         \midrule
        DSSM \cite{huang2013learning} & 0.6347 & 0.2886 \\ 
        % \midrule
        SBERT \cite{reimers-gurevych-2019-sentence} +Finetune & 0.7188 & 0.4235 \\
        MASM+Finetune & 0.6503 & 0.3238 \\
        % \midrule
        MASM+Click & 0.6150 & 0.2862 \\
        % \midrule
        MASM+LWR & 0.7751 & 0.4423 \\
        % \midrule
        MASM+LWR+Finetune & \ \ \textbf{0.7948*} & \ \ \textbf{0.4842*} \\
        \bottomrule
    \end{tabular}}
    \caption{Comparison results on our annotation dataset. * indicates the improvement over others are all statistically significant (p<0.01). }
    \label{main1}
\end{table}

\begin{table}[t]
    \centering
    % \begin{tabular}{p{4cm}|p{4cm}<{\centering}|cc}
    \scalebox{0.95}{
    % \begin{tabular}{l|m{1.5cm}<{\centering}|m{1cm}<{\centering}c}
    \begin{tabular}{l|m{2cm}<{\centering}|m{2cm}<{\centering}}
         \toprule
         \bf Model & \bf ROC-AUC & \bf Neg PR-AUC \\
         \midrule
         \midrule
        BM25 Sup (Rank+Emb) \cite{10.1145/3077136.3080832} & 0.6814 & 0.4802 \\
        Co-train \cite{10.1145/3289600.3291039} & 0.7746 & 0.5739 \\
        % \midrule
        MASM+Click & 0.6386 & 0.3682 \\
        MASM+LWR &  0.7827 & 0.5671 \\
        MASM+LWR+Finetune & \ \ \textbf{0.8047*} & \ \ \textbf{0.6012*} \\
        \bottomrule
    \end{tabular}
    }
    \caption{Comparison results on the test set of \citeauthor{10.1145/3289600.3291039}~\shortcite{10.1145/3289600.3291039}. * indicates the improvement over others are all statistically significant (p<0.01).}
    \label{main2}
\end{table}
 
\subsection{Results}
Table \ref{main1} presents the results of our model and baselines on the test set of Taobao annotation data described in Table \ref{human_data}. {\em DSSM} is widely used as a baseline model due to its simplicity and robust performance, but the results are far from satisfactory, which demonstrates the difficulty of the e-commerce relevance task. {\em MASM+Click} refers to our proposed model architecture but is trained on the original click-through data in a pair-wise fashion. We can see this model does not perform well. It is commonly observed that in the e-commerce relevance scenario pair-wise training from click-through data can bring benefits to model performance, compared to the point-wise training which fails to converge. But the performance is badly limited by the noisy click-through data. 
{\em MASM+Finetune} studies the most straightforward way to deal with the task, which uses gold-standard data to train the deep neural network model. With considerable amount of training data, the model can achieve better results than {\em MASM+Click} that does not use annotated training data. Unfortunately, the amount of data is still limited and the model is easy to overfit, considering the complicated e-commerce relevance task that deals with billions of queries and products. 
{\em SBERT} employs the pre-trained language model from large corpus and is fine-tuned on annotation data. As shown in Table \ref{main1}, {\em SBERT} performs much better and outperforms the {\em MASM+Click} baseline by a large margin.
These results indicate that introducing the pre-trained language model's knowledge is helpful for semantic matching tasks. Compared to the {\em MASM+Click} baseline which has the same model architecture, our {\em MASM+LWR} model with the proposed data construction method as well as new point-wise training objective gains substantial improvement of performance. Meanwhile, without the use of annotation data, {\em MASM+LWR} already outperforms the strong baseline {\em SBERT} which is pre-trained from a large corpus and fine-tuned with annotation data. With the proposed {\em LWR} dataset for training and human annotation dataset for fine-tuning, {\em MASM+LWR+Finetune} significantly outperforms all the baselines and achieves the state-of-the-art results.

Similar results are also observed on the test set of \citeauthor{10.1145/3289600.3291039}~\shortcite{10.1145/3289600.3291039}. Note that since we do not have the training set, we directly evaluate our models {\em MASM+Click}, {\em MASM+LWR}, and {\em MASM+LWR+Finetune} on this test set. The scores of {\em BM25 Sup} and {\em Co-training} are provided by \citeauthor{10.1145/3289600.3291039}~\shortcite{10.1145/3289600.3291039} instead of re-implementing them by ourselves. As we can see, {\em MASM+Click} does not perform well, while {\em BM25 Sup} can achieve better results by constructing more reasonable training labels instead of click or not. Surprisingly, {\em MASM+LWR} still outperforms {\em BM25 Sup} \cite{10.1145/3077136.3080832}, and is comparable with the state-of-the-art baseline {\em Co-training} \cite{10.1145/3289600.3291039} which uses annotation data to train. This performance demonstrates the effectiveness of our proposed data construction method as well as the well-designed loss function. When adding annotation data to fine-tune, the {\em MASM+LWR+Finetune} model can further improve the ROC-AUC results for about $2$ points and the Neg PR-AUC results for more than $3$ points, which outperforms  {\em Co-training} by a wide margin and achieves new state-of-the-art results. 

\subsection{Ablation Study}
We conduct ablation study to verify the effectiveness of the proposed \textit{LWR} dataset. By removing each type of data in Table~\ref{clickthrough_data} from the \textit{LWR} dataset, we compare the performance of \textit{MASM+LWR} on the test set in Table~\ref{human_data} to see the contribution of each type of click-through samples. Experimental results are presented in Table~\ref{ablation}. It can be seen that each type of samples contributes to the overall performance of \textit{MASM+LWR} differently. For the three types of relevant samples, \textit{Strong relevant} contributes the most, while surprisingly \textit{Relevant} contributes the least although this type of data accounts for the largest proportion. Considering the small size of \textit{Weak relevant} but also has significant effect to the overall performance, it indicates the model will benefit from learning to distinguish these difficult relevant samples. Generally, removing each type of relevant samples does not cause severe degradation of performance. Considering the data distribution will be significantly changed, it verifies our method is quite robust to the data distribution and the data construction of relevant samples. However, removing the irrelevant samples will cause significant drop of the performance, especially the \textit{Weak irrelevant} data. Without the \textit{Strong irrelevant} data, the model has noticeable loss of performance, which indicates introducing the random samples is an easy and cheap way to improve the relevance model. The \textit{Weak irrelevant} has much more influence, showing the challenge for the online e-commerce relevance model, which usually has to deal with hard-to-classify instances. Thus, the \textit{Weak irrelevant} samples introduced can effectively improve the model performance in this case.

\begin{table}[t]
    \centering
    % \begin{tabular}{p{4cm}|p{4cm}<{\centering}|cc}
    \scalebox{1.0}{
    % \begin{tabular}{l|m{1.5cm}<{\centering}|m{1cm}<{\centering}c}
    \begin{tabular}{l|m{2cm}<{\centering}|m{2cm}<{\centering}}
         \toprule
         \bf Model & \bf ROC-AUC & \bf Neg PR-AUC \\
         \midrule
         \midrule
        MASM+LWR & \textbf{0.7751} & \textbf{0.4423} \\ 
        % \midrule
        \ w/o Strong relevant & 0.7593 & 0.4231 \\
        \ w/o Relevant & 0.7669 & 0.4399 \\
        % \midrule
        \ w/o Weak relevant & 0.7632 & 0.4269 \\
        % \midrule
        \ w/o Weak irrelevant & 0.6492 & 0.3552 \\
        % \midrule
        \ w/o Strong irrelevant & 0.7423 & 0.3859 \\
        % \midrule
        \bottomrule
    \end{tabular}}
    \caption{Comparison results of removing each type of data in the LWR dataset.}
    \label{ablation}
\end{table}

\subsection{Score Distribution}
\begin{figure}
    \centering
    \includegraphics[width=0.99\columnwidth]{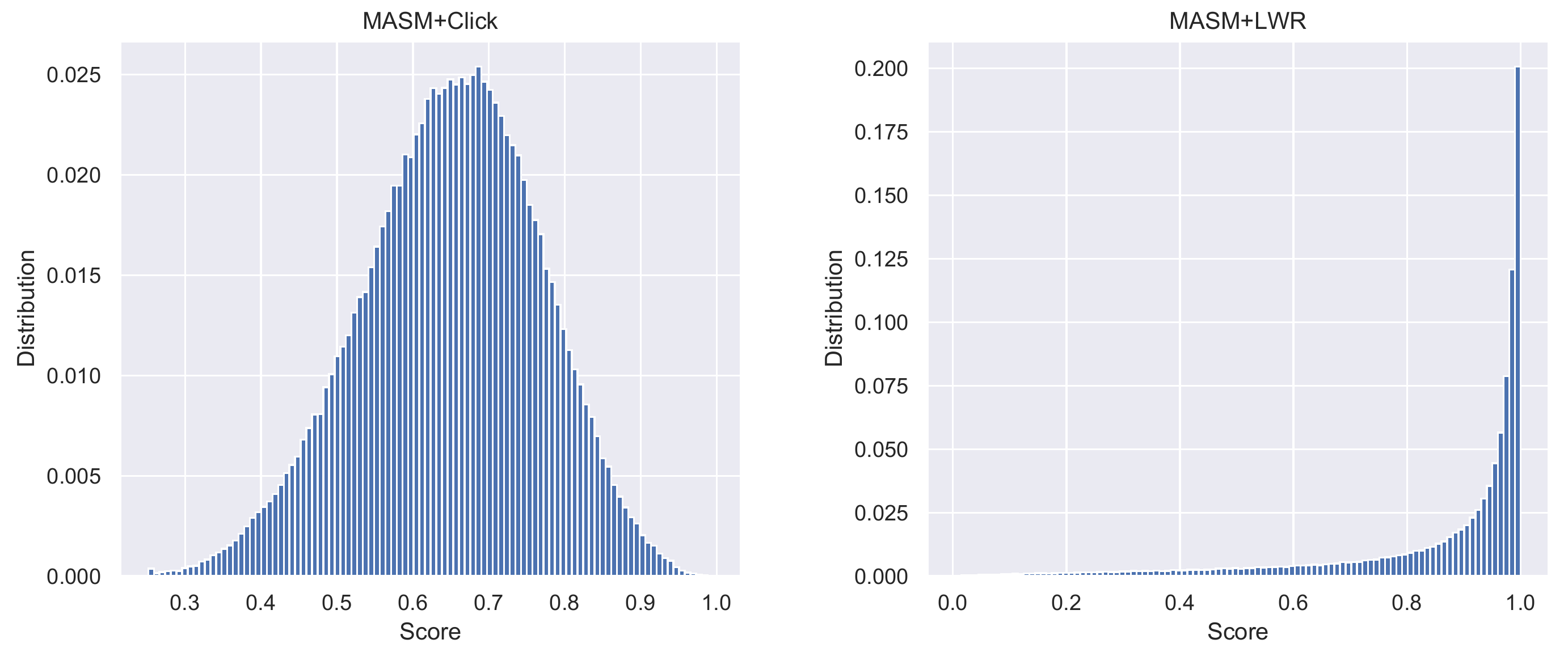}
    \caption{The score distribution of MASM trained with different datasets.}
    \label{score_dist}
\end{figure}

To analyze the effect of our proposed training paradigm, we compare the distribution of predicted relevance scores of different models on the test set in Table \ref{human_data}. Figure \ref{score_dist} presents the model trained with the click-through data in a pair-wise fashion (\textit{MASM+Click}) and the model trained with our constructed dataset (\textit{MASM+LWR}). It can be found that the predicted scores of a typical pair-wise model trained with click data roughly obey normal distribution. However, about $80$\% of the test data are positive samples, but the predicted scores of \textit{MASM+Click} rarely approach to $1$. On the contrary, \textit{MASM+LWR} predicts most test samples to high scores. This verifies the proposed fine-grained data construction method and loss function can learn a robust model with desirable score distribution, which is close to the actual situation and is more appropriate for online serving.

\subsection{Online Evaluation}
The proposed model has been deployed online in the search engine of Taobao. The initial online serving latency is about 50ms, which is quite computational expensive due to Taobao's extremely large traffic. Fortunately, the proposed model has the advantage of representation-based architecture, i.e. the multi-aspect query vectors and product vectors can be pre-computed. Then, online serving will only need to compute the lightweight interaction and MLP layers. After this optimization, the online latency drops to 8ms only.

Online A/B testing is conducted to verify the effectiveness of the proposed model. The baseline experiment is the previous online relevance model, which is an in-house version of \textit{Co-training} \cite{10.1145/3289600.3291039}. The test experiment is to replace the \textit{Co-training} model with the \textit{MASM+LWR+Finetune} model. Both experiments take about $2$\% proportion of Taobao search traffic. The experiments last for $7$ days. On average, the proposed model improves Gross Merchandise Volume for about $0.55$\%. Human annotation results show that the proposed model can improve the rate of \textit{Relevant} by $0.76$\% points. Online A/B testing verifies the proposed model is superior to previous state-of-the-art models, and can achieve significant online profit considering the extremely large traffic of Taobao everyday. The proposed model has already been deployed to serve the entire Taobao search traffic for over a year.

\section{Conclusion and Future Work}
In this paper, we study an industrial task of measuring the semantic relevance for queries and products in e-commerce. The problem is challenging due to the lack of high-quality training data, and the requirement for computationally efficient model. We propose a multi-aspect attention technique to promote the interaction of representation-based model. To train the relevance model, we propose a novel way to construct a fine-grained dataset from the click-through data and come up with a new point-wise training objective. This strategy allows the model to learn the relative relevance between different training instances similarly as pair-wise architecture, but uses a point-wise loss function which greatly reduces the training costs. Our proposed training paradigm can maintain the advantages of pair-wise training in a point-wise fashion and is able to learn a robust relevance model with desirable score distribution. The proposed model achieves promising results on both offline and online experiments, and has been deployed to serve the entire Taobao search traffic for over a year, which improves both semantic relevance and Gross Merchandise Volume.

The dataset and training objective proposed in this paper are independent of the model architecture. Therefore, in our future work, we will explore the application on other models, e.g. more powerful interaction-based models or the BERT model. How to make these complicated models able to be deployed online is also our future direction.

\bibliographystyle{ACM-Reference-Format}
\bibliography{main}


\end{document}